\documentclass[reqno,11pt]{amsart}

\usepackage{fancyhdr,bbm}
\usepackage{datetime}
\settimeformat{ampmtime}
\usepackage[svgnames]{xcolor}
\usepackage{setspace,palatino}
\usepackage{pgf}
\usepackage{tikz}
\usepackage[scaled]{helvet} 
\usepackage{courier} 
\normalfont
\usepackage[T1]{fontenc}
\usepackage{booktabs}

\usetikzlibrary{patterns}
\usepackage{soul}
\usepackage{mathrsfs}
\usepackage{stmaryrd}
\usepackage[hidelinks=true]{hyperref}
\usepackage{natbib}
\usepackage{xfrac}
\usepackage[color=Orange]{todonotes}
\usepackage{layout}
\usepackage{microtype}
\usepackage{mathtools}
\usepackage{multirow}
\usepackage{floatrow}
\usepackage{subfig}
\usepackage[multiple]{footmisc}

\newtheorem{lemma}{Lemma}

\newtheorem{theorem}{Theorem}

\usepackage{cleveref}

\crefname{figure}{figure}{figures}
\creflabelformat{figure}{#2#1#3}
\crefname{equation}{equation}{equations}
\creflabelformat{equation}{#2(#1)#3}
\crefname{lemma}{lemma}{lemmas}
\creflabelformat{lemma}{#2#1#3}
\crefname{theorem}{theorem}{theorems}
\creflabelformat{lemma}{#2#1#3}
\crefname{condition}{condition}{conditions}
\creflabelformat{condition}{#2#1#3}
\crefname{assumption}{assumption}{assumptions}
\creflabelformat{assumption}{#2#1#3}
\crefname{appendix}{appendix}{appendices}
\creflabelformat{appendix}{#2#1#3}
\crefname{enumi}{}{}
\creflabelformat{enumi}{(#2#1#3)}

\newtheorem{assumption}{Assumption}

\DeclareMathOperator{\argmax}{argmax}

\usepackage{amssymb}

\renewcommand{\Pr}{\mathbb{P}}

\newcommand\abstraction{We consider the estimation of dynamic discrete choice models in a semiparametric setting, in which the per-period utility functions are known up to a finite number of parameters, but the distribution of utility shocks is left unspecified.   This semiparametric setup differs from most of the existing identification and estimation literature for dynamic discrete choice models.   To show identification we derive and exploit a new Bellman-like recursive representation for the unknown quantile function of the utility shocks.   Our estimators are straightforward to compute; some are simple and require no iteration, and resemble classic estimators from the literature on semiparametric regression and average derivative estimation.   Monte Carlo simulations demonstrate that our estimator performs well in small samples. To highlight features of this estimator, we estimate a structural model of dynamic labor supply for New York City taxicab drivers.
 \\ 

\noindent
\textbf{Keywords:} Semiparametric estimation, Dynamic discrete choice model, Average derivative estimation, Taxicab industry, Labor supply\\

\noindent
\textbf{JEL}: C14, D91, C41, L91
}

\usepackage{subfig}

\begin{document}

\thispagestyle{empty}

\title{Semiparametric Estimation of Dynamic Discrete Choice Models}

\thanks{$^*$We thank  Hassan Afrouzi, Saroj Bhattarai, Kirill Pogorelskiy, Eduardo Souza-Rodrigues, Sorawoot (Tang) Srisuma, and seminar
  participants at the University of Arizona, University of Washington (Seattle), UT Austin and Texas Metrics Camp for helpful discussion. }
\author{
\href{mailto: nibu@utexas.edu}{Nicholas Buchholz$^{\star}$}}
\thanks{$^{\star}$Department of Economics, University of Texas at Austin, \href{mailto: nibu@utexas.edu}{ nibu@utexas.edu}}

\author{
\href{mailto:mshum@caltech.edu}{Matthew Shum$^\dag$}}
\thanks{$^\dag$Division of Humanities and Social Sciences,  California Institute of Technology,
\href{mailto:mshum@caltech.edu}{mshum@caltech.edu}}
\author{
\href{mailto:h.xu@austin.utexas.edu}{Haiqing Xu$^{\ddag}$}}
\thanks{$^{\ddag}$Department of Economics, University of Texas at Austin, \href{mailto:h.xu@austin.utexas.edu}{h.xu@austin.utexas.edu}}

\date{\today}

\maketitle

\begin{abstract}
 \abstraction
\end{abstract}

\vspace{5ex}

\clearpage

\section{Introduction}

 The dynamic discrete choice (DDC) framework, pioneered by \cite{wolpin1984estimable}, \cite{pakes_patents}, \cite{rust1987optimal,rust1994structural}, has gradually become the workhorse model for modelling dynamic decision processes in structural econometrics.   Such models, which can be considered  an extension of McFadden's \citeyear{mcfadden1978modelling,mcfadden1980econometric} classic random utility model to a dynamic decision setting, have been used to model a variety of economic phenomenon is ranging from labor and health economics to industrial organization, public finance, and political economy.  More recently, the DDC framework has also been the starting point for the empirical dynamic games literature in industrial organization.

 In this paper, we consider identification and estimation of a class of semiparametric DDC models, in which the utility indices are parametrically specified, but the shock distribution is left unspecified.   Since the utility shocks are typically interpreted as idiosyncratic and unpredictable shocks to preferences which cause agents' choices to vary over time even under largely unchanging economic environments, it is reasonable to leave their distribution unspecified.   We study conditions under which the model structure (consisting of the finite-dimensional parameters in the utility indices, and the infinite-dimensional nonparametric shock distribution) is identified.  Our identification argument is constructive, and we propose an estimator based upon it.   

The semiparametric DDC framework which we focus on in this paper is novel relative to most of the existing literature on the identification and estimation of DDC models, which considers the case where the utility shocks are fully (parametrically) specified.   This reflects an important result in \cite{magnac2002identifying}, who argue that in these models, the single-period utility indices for the choices are (nonparametrically) identifiable only when the distribution of the utility shocks is completely specified.   Based on Magnac--Thesmar's ``impossibility'' result, many recent estimators for and applications of DDC models have considered a structure in which the single-period utility indices are left unspecified, but the utility shock distribution is fully specified (and usually logistic, leading to the convenient multinomial logit choice probabilities).

For identification, we derive a new recursive representation for the unknown quantile function of the utility shocks.  Accordingly, we obtain a single-index representation for the conditional choice probabilities in the model, which permits us to estimate the model using classic estimators from the existing semiparametric binary choice model literature.  Specifically,  we use Powell, Stock and Stoker's (\citeyear{powell1989semiparametric},PSS) kernel-based estimator to estimate the the dynamic discrete choice model.   Our estimator has the same asymptotic properties as PSS's original estimator (for static discrete-choice models), under under additional mild conditions.  Moreover, the estimator is computationally simple and non-iterative.     Monte Carlo simulations demonstrate that our estimator performs well in small samples. To highlight features of this estimator, we estimate a structural model of dynamic labor supply for New York City taxicab drivers.

\subsection{Literature}
 This paper builds upon several strands in the existing literature.
 The semiparametric binary choice literature (e.g. \cite{manski1975maximum,manski1985semiparametric}, \cite{powell1989semiparametric}, \cite{ichimura1991semiparametric}, \cite{horowitz1992smoothed}, \cite{klein1993efficient}, and \cite{lewbel1998semiparametric}, among many others) is an important antecedent.   However, an important substantive difference is that, because these papers focus on a static model, the shock distribution is treated as a nuisance element, and estimation of it is not considered.  In contrast, with a dynamic model, the shock distribution must necessarily be estimated, because it affects the beliefs that decision makers have regarding their future payoffs.   Hence, the need to estimate both the utility parameters as well as the shock distribution represents an important point of divergence between our paper and the previous semiparametric discrete choice literature; nevertheless, as we will point out, the estimators we propose take a form which is very similar to the estimators in these papers.

To our knowledge, only \cite{norets2014semiparametric}, \cite{Chen2014}, and
 \cite{Blevins2014} consider identification of dynamic models in which the
 error distribution is left unspecified.  \cite{norets2014semiparametric} focus on the
 discrete state case, and derive (joint) bounds on the error distribution
 and per-period utilities which are consistent with an observed vector of
 choice probabilities.   We consider the case with continuous state
 variables, and discuss nonparametric identification and estimation.  With
 a discrete state space, there can never be point identification when the
 error distribution has continuous support.   When the state space is
 continuous, however, point identification is possible under  some support conditions and a location--scale normalization on the error distribution, as we show.   

\cite{Chen2014} considers the identification of dynamic models, and, as we do here, obtains simple estimators for the model parameters, similar to familiar estimators in the semiparametric
discrete choice literature.   His approach exploits exclusion restrictions (that is, that a subset of the state variables affect only current utility, but not agents' beliefs about future utilities).  In contrast, we do not use exclusion restrictions, but rather exploit the optimality conditions to derive  a new recursive representation of the quantile function for the unobservables which allows us to identify and estimate the model.
\cite{Blevins2014} considers a very general class of dynamic models in
which agents can make both discrete and continuous choices, and the shock
distribution can depend on some of the state variables.   Under exclusion
restrictions, he shows the nonparametric identification of both the
per-period utility functions, as well as the error distribution.  

 Another important related paper is \cite{srisuma2012semiparametric}, who pioneered the use of tools for solving type 2 integral equations for estimating dynamic discrete-choice models.  We show that, besides the Bellman equation, other structural relations in the dynamic model also take the form of type 2 integral equations.
In particular, when viewed as a function of the choice probability, the
(unknown) quantile function for the utility shocks can also be recursively
characterized via a Bellman-type equation, and hence methods for solving
for the value function in the ``usual'' Bellman equation (either value
function iteration or ``forward simulation'') can also be applied to
solving for this quantile function.

\section{Single Agent Dynamic Discrete Choices Model}
Following \cite{rust1987optimal}, we consider a single--agent infinite-horizon binary decision problem. At each time period $t$, the agent observes state variables $X_t\in \mathbb R^k$, and choose a binary decision $Y_{t}\in\{0,1\}$ to maximize her expected utility. The per--period utility is given by 
$$
u_{t}(Y_t,X_{t},\epsilon_{t})=\left\{\begin{array}{cl}   W_{1}(X_{t})^\intercal\theta_1 +\epsilon_{1t}, & \text{ if }Y_{t}=1;\\
W_{0}(X_{t})^\intercal\theta_0+ \epsilon_{0t}, & \text{ if }Y_{t}=0.
\end{array}\right.
$$   
In the above $W_{0}(X_{t})\in\mathbb R^{k_0}$ (resp. $W_{1}(X_{t})\in\mathbb R^{k_1}$) denote known transformations of the state variables $X_{t}$ which affect the per--period utility from choosing $Y_{t}=0$ (resp. $Y_{t}=1$), and $\epsilon_t\equiv(\epsilon_{0t},\epsilon_{1t})^\intercal\in\mathbb R^2$ are the agent's action specific payoff shocks;  For $d=0,1$, $\theta_d\in\mathbb R^{k_d}$ are structural parameters of our interest. This specification of the per-period utility functions as single-indices of the transformed state variables $W_{0}(X)$ and $W_{1}(X)$ encompasses a majority of the existing applications of dynamic discrete-choice models, and hence poses little loss in generality.  The utility of action $0$ is not normalized to be zero for the reasons discussed in \cite{norets2014semiparametric}. 

Moreover, let $\beta\in(0,1)$ be the discount factor and $f_{X_{t+1},\epsilon_{t+1}|X_{t},\epsilon_t,Y_t}$ be the Markov transition probability density function that depends on the state variable as well as the decision. For notational simplicity, we will use the shorthand $W_d$ for $W_d(X)$ ($d=0,1$) and suppress the explicit dependence upon the state variables $X$.

The agent maximizes the expected discounted sum of the per-period payoffs: 
$$
\max_{\left\{y_{t}, y_{t+1},\ldots\right\}} \mathbb{E} \left\{\sum_{s=t}^{\infty} \beta^{s-t} u_s(y_{s},
X_{s},\epsilon_{s})|X_{t},\epsilon_{t}\right\},\quad \text{s.t.}\quad f_{X_{s+1},\epsilon_{s+1}|X_{s},\epsilon_s,Y_{s}}.
$$ We assume stationarity of the problem, which implies that the problem is invariant to the period  $t$.   Because of this, we can omit the $t$ subscripts, and we use  primes (${}'$) to denote next period values. Let  $V(X,\epsilon)$ be the value function given $X$ and $\epsilon$. By Bellman's equation, the value function can be written as
\begin{equation*}
V(X,\epsilon) = \text{max}_{y\in\{0,1\}} \left\{\mathbb E [u(y,X, \epsilon)|X,\epsilon] + \beta \mathbb{E} [V(X',\epsilon')|X,\epsilon,Y=y]\right\},
\end{equation*}
and then the agent's optimal decision is given by
$$
Y = \argmax_{y\in \{0,1\}} \left\{\mathbb E [u(y,X, \epsilon)|X,\epsilon] + \beta \mathbb{E} [V(X',\epsilon')|X,\epsilon,Y=y]\right\}.
$$
Unlike much of the existing literature, we do not assume the distribution of the utility shocks $(\epsilon_{0t}, \epsilon_{1t})$ to be known, but  treat their distribution as a nuisance element for the estimation of $\theta$.   In a static setting, such flexibility may not be necessary as a flexible specification of $u(X,Y)$ may be able to accommodate any observed pattern in the choice probabilities even when the distribution of utility shocks is parametric.\footnote{\cite{mcfadden_train2000} show such properties for the mixed logit specification of static multinomial choice models.}  However, in a dynamic setting, the distribution of utility shocks also play the role of agents' beliefs about the future evolution of state variables (i.e. they are a component in the transition probabilities $f_{X', \epsilon'|X,\epsilon,Y}$) and hence parametric assumptions on this distribution are not innocuous.

\subsection{Characterization of the value function}
In this subsection, we characterize the value function $V(X,\epsilon)$ and  the expected value function given $X$, i.e., $V^e(X)\equiv \mathbb E [V(X,\epsilon)|X]$. Both value functions are useful to characterize the equilibrium in our dynamic model.  Let $F_A$ and $F_{A|B}$ denote the CDF and the conditional CDF for generic random variables $A$ and $B$, respectively. 
\begin{assumption}[Conditional Independence Assumption]
\label{as1}
The transition density satisfies the following condition: $F_{X',\epsilon'|X,\epsilon,Y}=F_{\epsilon'}\times F_{X'|X,Y}$. Moreover, $F_{\epsilon'}=F_{\epsilon}$.
\end{assumption}
\noindent
\Cref{as1} is strong but fundamental in the current literature; 
our approach requires no additional exclusion restrictions beyond those typically assumed in the parametric dynamic discrete-choice literature (see e.g.  \cite{rust1987optimal} and \cite{srisuma2012semiparametric}).\footnote{On the other hand, as in \cite{Blevins2014}, it is possible to relax the independence assumption to one where the state variables can be divided into two groups $X=(X_A, X_B)$ such that $\epsilon\perp X_B|X_A$ ($\epsilon$ is independent of $X_B$ given $X_A$).   It appears the identification and estimation procedure described in this paper follow through, with the additional conditioning on $X_A$ at every step.  
} 

Under \cref{as1}, the value function can be written as
\begin{equation*}
V(X,\epsilon) 
= \max\Big\{ W_1^\intercal\theta_1+\epsilon_1
+ \beta \mathbb{E} [V(X',\epsilon')|X,Y=1],\
W_0^\intercal\theta_0+\epsilon_0+ \beta \mathbb{E} [V(X',\epsilon')|X,Y=0]\Big\}.
\end{equation*}
Let $\eta=\epsilon_0-\epsilon_1$. Then, the equilibrium decision  maximizing the value function can be written as
\[
Y=\mathbb 1 \{\eta\leq \eta^*(X)\}.
\] where the cutoff $\eta^*(X)$ is defined as
\begin{equation}
\label{cutoff}
\eta^*(X)\equiv W_1^\intercal\theta_1-W_0^\intercal\theta_0+ \beta \left\{\mathbb{E} [V(X',\epsilon')|X,Y=1]- \mathbb{E} [V(X',\epsilon')|X,Y=0]\right\}.
\end{equation}

Moreover, let $  u^e(X)$ be the expected per--period utility conditional on $X$, i.e.,
\begin{equation}
\label{ustar}
  u^e(X)  \equiv  \mathbb E (\epsilon_{0})
  +W_1^\intercal\theta_1\cdot F_\eta\big(\eta^*(X)\big)+W_{0}^\intercal\theta_0\cdot [1-F_\eta\big(\eta^*(X)\big)]-\mathbb E\big\{\eta\cdot \mathbb 1[\eta\leq\eta^*(X)]\big\},
  \end{equation} where $F_\eta$ is the CDF of $\eta$. Thus, the Bellman equation can be rewritten as 
  \begin{equation}
  \label{bellman}
  V^e(X)=u^e(X)+\beta\cdot \mathbb E [V^e(X')|X].
  \end{equation} It is worth pointing out that eq. \eqref{bellman} is essentially a Fredholm Integral Equation of the second kind  (FIE--2); See e.g. \cite{zemyan2012classical}. Essentially, FIE--2 is a linear equation system in functional space, which is well--known to have a unique analytic solution under some sufficient and necessary conditions. 
 
   \begin{assumption}
 \label{as_regular}
 For all $s\geq 1$, we have $\mathbb E \big(\|W_d^{[s]}\||X\big)<\infty$, where $( ^{[s]})$ denotes the next $s$ period values.
\end{assumption}
\noindent \Cref{as_regular} holds when $W_d(\cdot)$ are bounded functions. 
  

Given these assumptions, the next lemma applies the Fredholm theorem to obtain a solution of the expected value function to the Bellman equation. (Similar results are utilized in \cite{srisuma2012semiparametric}.)

 \begin{lemma}
\label{lemma1}
Suppose \cref{as1,as_regular} hold. Then, we have 
\begin{equation}
\label{valuef}
V^e(x)=u^e(x)+  \beta\int_{\mathscr S_X} R^*(x', x;\beta)\cdot u^e( x')d x',\ \ \forall x\in\mathscr S_X,
\end{equation}
where  $R^*(x',x;\beta) = \sum_{s=1}^{\infty} \beta^{s-1} f_{X^{[s]}|X}(x'|x)$ is the resolvent kernel generated by the FIE  eq. \eqref{bellman}.
\end{lemma}
\noindent
More succinctly, eq. \eqref{valuef} can be rewritten as
\begin{equation}
\label{eq2}
V^e(X)=u^e(X)+ \sum_{s=1}^{\infty}\beta^{s}\cdot \mathbb E [u^e(X^{[s]})|X].
\end{equation} In operator notation, eq. \eqref{eq2} denotes exactly the ``forward integration'' representation of the value function, which is familiar from many two-step procedures for estimating dynamic discrete choice models \citep[see e.g.][]{hotz1993conditional,bajari2007estimating,hong2010pairwise}.   In the special case when the state variables $X$ are finite and discrete-valued (taking $k<\infty$ values), the Bellman equation is a system of linear equations which can be solved for the value function \citep[cf.][]{aguirregabiria2007sequential,pesendorfer2008asymptotic} and in that case, the resolvent kernel is just the inverse matrix $(I-\beta F_{X'|X})^{-1}$ where $F_{X'|X}$ denotes the $k\times k$ transition matrix for $X$.

\subsection{Equilibrium Condition} To characterize the equilibrium, the key of our approach is to solve for the cutoff value $\eta^*$ that depends on the state variables $X$ (through the transformations $W_1(X)$ and $W_0(X)$). By using eq. \eqref{eq2}, along with \Cref{lemma1}, eq. \eqref{cutoff} becomes 
\begin{equation}
 \eta^*(X)=W_1^\intercal\theta_1-W_0^\intercal\theta_0+ \sum_{s=1}^{\infty}\beta^{s}\left\{\mathbb E [u^e(X^{[s]})|X,Y=1]- \mathbb E [u^e(X^{[s]})|X,Y=0]\right\}.
\end{equation} Moreover, let $\phi_d(X)
\equiv (-1)^{d+1}W_{d}
+ \sum_{s=1}^\infty\beta^s\Big\{\mathbb E \big[W_d^{[s]}\mathbb 1_{Y^{[s]}=d}|X,Y=1\big]-\mathbb E \big[W_d^{[s]}\mathbb 1_{Y^{[s]}=d}|X,Y=0\big]\Big\}$. Then,  it follows from  \eqref{ustar}, 
\begin{multline}
\label{eq4}
\eta^*(X)=\phi^\intercal (X)\cdot\theta\\
- \sum_{s=1}^{\infty}\beta^{s}\left\{\mathbb E \big[\eta^{[s]}\mathbb 1(\eta^{[s]}\leq\eta^*\big(X^{[s]}\big))\big|X,Y=1\big]-\mathbb E \big[\eta^{[s]}\mathbb 1(\eta^{[s]}\leq\eta^*\big(X^{[s]}\big) )\big|X,Y=0\big]\right\},
\end{multline} where $\phi(X)=(\phi_0^\intercal(X),\phi_1^\intercal(X))^\intercal$ and $\theta= (\theta_0^\intercal,\theta_1^\intercal)^\intercal$.

Eq. \eqref{eq4} characterizes the equilibrium decision rule in the single--agent infinite-horizon binary decision problem.  Alternatively, we can rewrite it  using a resolvent kernel:
\begin{equation*}
\eta^*(x)=\phi^\intercal (x)\cdot\theta
- \beta\int_{\mathscr S_X}\mathbb E[\eta'\cdot \mathbb 1(\eta'\leq\eta^*(x')]\cdot g(x',x;\beta) dx',\ \ \ \forall x\in\mathscr S_X,
\end{equation*} where $g(x',x;\beta)= \sum_{s=1}^\infty \beta^{s-1} [f_{X^{[s]}|X,Y}(x'|x,1)- f_{X^{[s]}|X,Y}(x'|x,0)]$.  Given the structural parameters $\theta_0$, $\theta_1$, $F_{\eta}$ and $f_{X'|X,Y}$, in principal one can solve the threshold $\eta^*(\cdot)$.\footnote{However, if one were to use this equation to solve for $\eta^*(\cdot)$ via simulation or computation, note that $g(x',x;\beta)$ also contains $\eta^*(\cdot)$ implicitly through the transition density $f_{X^{[s]}|X,Y}(\cdot|\cdot,\cdot)$.}

\section{Identification}
Next, we develop an identification strategy that does not involve solving the Markov equilibrium in the dynamic decision problem. To clarify ideas,  we first provide identification of structural parameter $\theta\in\Theta\subseteq \mathbb R^{k_\theta}$  (where $k_\theta\equiv k_0+k_1$) in a fully parametric model, i.e., assuming $F_\eta$ is known. Then, we establish semiparametric identification of our model by a two--step approach: we first identify $F_\eta$ up to the finite dimensional parameter $\theta$. In the second step, we represent the agent's choice by a single--index representation. Therefore, the identification of $\theta$ follows the literature. 

A key feature in our semiparametric identification is that we require (at least) one argument in the state variables $X_t$ to have continuous variation, which is also the case in the semiparametric identification of the single--index binary response model in the static setting. See e.g.  \cite{manski1975maximum}. Moreover, we show that the quantile function of $F_\eta$ is identified on the support of the agent's choice probabilities under a location--scale normalization. This result also corresponds to the findings in static binary response models. 
\subsection{Intermediate step: Parametric Identification} As a building block towards the more general semiparametric results below, we first consider parametric identification of the model, assuming that the researcher know $F_{\eta}$, the distribution function for the utility shocks.   Parametric identification in this setting (with continuous state variables)  has also been established previously in  \cite{srisuma2012semiparametric}. Our identification of $\theta$ is constructive, which has an single index structure and suggests an OLS estimator.  To begin with, we introduce the following assumption. 
\begin{assumption}
\label{as2}
Let $\eta$ be continuously distributed with the full support $\mathbb R$.
\end{assumption}
\noindent
\Cref{as2} is a weak condition widely used in semiparametric binary response models \citep[see e.g.][]{horowitz2009semiparametric}.  Under \cref{as2},  $F_\eta$  is strictly increasing  on its support $\mathbb R$. Let $Q$ be the quantile function of $F_\eta$, i.e., $Q=F^{-1}_\eta$. 

Let  $p(x)=\Pr(Y=1|X=x)$, which obtains directly from the data.  Under \cref{as2}, $0<p(x)<1$ for all $x\in\mathscr S_X$ and $\eta^*(x)=Q(p(x))$. 
Moreover, using the substitution $\tau\rightarrow Q(\tau)$, then we have
\[
\mathbb E[\eta\cdot \mathbb 1(\eta\leq Q(p)]=\int \tau\cdot \mathbb 1 (\tau\leq Q(p)) d F_\eta(\tau)=\int_0^pQ (\tau)d\tau.
\] 
From above discussion, it is straightforward that we obtain the following lemma.
\begin{lemma}
\label{lemma2}
Suppose \cref{as1,as_regular,as2,} hold. Then  we have
\begin{multline}
\label{eq6}
Q(p(X))+\sum_{s=1}^\infty\beta^{s}\left\{\mathbb E \Big[\int_0^{p(X^{[s]})} Q(\tau)d\tau\big|X,Y=1\Big]-\mathbb E \Big[\int_0^{p(X^{[s]})} Q(\tau)d\tau\big|X,Y=0\Big]\right\}\\
=\phi^\intercal(X)\cdot \theta.
\end{multline}
\end{lemma}
\noindent
As a matter of fact, eq. \eqref{eq6} is the key restriction for our identification and estimation analysis, where  the number of restrictions equals to the size of the support $\mathscr S_X$.

When $Q_\eta$ is given, then everything in \eqref{eq6} is known except for $\theta$.  If, in addition, the matrix $\mathbb E [\phi(X)\phi^\intercal (X)]$ is invertible, then $\theta$ can be estimated using nonlinear least-squares on eq. \eqref{eq6}.  This approach is related to \cite{pesendorfer2008asymptotic}. The full rank of $\mathbb E [\phi(X)\phi^\intercal (X)]$ requires that if the transformed state variables $W_0(X)$ and $W_1(X)$ contain a common components $W_c(X)$, then there is $\mathbb E [W_c(X')|X,Y=0]\neq \mathbb E [W_c(X')|X,Y=1]$. Such a necessary condition rules out the case that variables without any dynamic transition (e.g. the constant) are included in both transformed state variables.\footnote{ In this case, it is also required that the discount rate $\beta\neq 0$, otherwise $\phi_0(X)=W_0(X)$ and $\phi_1(X)=W_1(X)$, which clearly invalidates the rank condition due to the common term $W_c(X)$.} 
  

%

\subsection{Semiparametric Identification}
Without making any distributional assumptions on $\eta$, we now discuss the identification of $\theta$ as well as $Q_\eta$. In what follows we will assume $W_0(X)$ or $W_1(X)$ has at least one continuous argument such that $p(X)$ is continuously distributed.   It is well--known that the continuity of covariates is crucial for the semiparametric identification in the static binary response model. This is still the case in our dynamic binary decision model. 

 Intuitively, the number of restrictions imposed by \eqref{eq6} depends on the richness of the support $\mathscr S_X$.  
 For identification of $Q_\eta$ (up to $\theta$), however we only exploit variations in the choice probabilities $p(X)$. For notational simplicity, we assume  the choice probability $p(X)$ is continuously distributed on a closed interval.\footnote{This interval--support restriction can be relaxed at expositional expense. For instance, suppose $\mathscr S_{p(X)}$ is a non--degenerate compact subset of $[0,1]$.  All our identification arguments below still hold by replacing the integral region $[\underline p, \ \overline p]$ with $\mathscr S_{p(X)}$.} 

\begin{assumption}
\label{as3}
(i) Let $p(X)$ be continuously distributed; (ii) let the support of $p(X)$ be a closed interval, i.e., $[\underline p, \overline p]\subseteq [0,1]$.
\end{assumption}
\noindent
In contrast, when $p(X)$ only has discrete variation (which typically arises when the state variables $X$ themselves have only discrete variation),  \cite{norets2014semiparametric} show that the distribution of $\eta$ is partially identified. 

For each $p\in [\underline p, \overline p]$, let  $z(p)=\mathbb E[\phi(X)|p(X)=p]$. We now take the conditional expectation given $p(X)=p$ on both sides of eq. \eqref{eq6}. By the law of iterated expectation,  we have
\begin{multline*}
Q(p)+\sum_{s=1}^\infty\beta^{s}\left\{\mathbb E \Big[\int_0^{p(X^{[s]})} Q(\tau)d\tau\big|p(X)=p,Y=1\Big]-\mathbb E \Big[\int_0^{p(X^{[s]})} Q(\tau)d\tau\big|p(X)=p,Y=0\Big]\right\}\\
=z(p)^\intercal\cdot\theta.
\end{multline*}
The above discussion is summarized by the following lemma.
\begin{lemma}
\label{lemma3}
Suppose \cref{as1,as_regular,as2,as3} hold. Then, we have
\begin{equation}
\label{eq7}
Q(p)+\beta\int_{\underline p}^{\overline p}\int_{\underline p}^{p'}Q(\tau)d\tau \cdot \pi(p',p;\beta) dp' 
=z(p)^\intercal \cdot \theta, \ \ \forall p\in [\underline p, \overline p].
\end{equation} where $\pi(p',p;\beta)= \sum_{s=1}^\infty \beta^{s-1} [f_{p(X^{[s]})|p(X),Y}(p'|p,1)- f_{p(X^{[s]})|p(X),Y}(p'|p,0)]$. 
\end{lemma}
\noindent
By definition, $\pi(p',p;\beta)$ is the difference of the discounted
aggregate densities of the future choice probabilities, conditional on the
current choice probability and (exogenously given) action, which can be obtained directly from the data. 

Note that eq. \eqref{eq7} is also an  FIE--2. To see this,   let $ 
\Pi(p',p;\beta)\equiv \sum_{s=1}^\infty \beta^{s-1} [F_{p(X^{[s]})|p(X),Y}(p'|p,1)- F_{p(X^{[s]})|p(X),Y}(p'|p,0)]$. Then, the second term of eq. \eqref{eq7} can be rewritten as
\begin{align}
\nonumber\int_{\underline p}^{\overline p}\int_0^{p'}Q(\tau)d\tau \cdot \pi(p',p;\beta) dp'&=\int_0^{1}Q(\tau)\cdot \int_{\underline p}^{\overline p}\mathbb 1(\tau\leq p') \cdot \pi(p',p;\beta) dp'd\tau\\
\nonumber &=-\int_0^{1}Q(\tau)\cdot\Big[ \int_{\underline p}^{\overline p}\mathbb 1(p'<\tau)\pi(p',p;\beta) dp'\Big] d\tau\\
\nonumber &=-\int_{\underline p}^{\overline p} Q(\tau)\cdot \Pi(\tau,p;\beta)d\tau,
\end{align} where the second step comes from the fact $\int_{\underline p}^{\overline p} \pi(p',p;\beta) dp'=0$ and the last step is because $\Pi(p',p,\beta)=0$ for all $p'\not\in [\underline p,\overline p]$.  Hence, we obtain the following FIE--2: 
\begin{equation*}
Q(p)-\beta \int_{\underline p}^{\overline p} Q(\tau)\cdot \Pi(\tau,p;\beta)d\tau
=z(p)^\intercal \cdot \theta, \ \ \forall p\in [\underline p, \overline p].
\end{equation*}
By solving this equation, we can identify $Q(\cdot)$ on $[\underline p, \overline p]$ up to the finite dimensional parameter $\theta$.  
\begin{assumption}
\label{as4}
Let $\beta^2\cdot \int_{\underline p}^{\overline p}\int_{\underline p}^{\overline p} \Pi^2(p', p;\beta) dp' dp<1$.
\end{assumption}
\noindent
\Cref{as4} is introduced for the uniqueness of the solution. This
assumption a high level condition but testable.

 \begin{lemma}
 \label{lemma4}
Suppose \cref{as1,as_regular,as2,as3,as4} hold. Then,  $Q_\eta$ is point identified on $[\underline p, \overline p]$ up to the finite dimensional parameter $\theta$: 
\begin{equation}
\label{eq8}
Q(p)=\left\{z(p)-\beta\int_{\underline p}^{\overline p}  R(p',p;\beta)\cdot z(p') dp'\right\}^\intercal\cdot \theta,\ \ \ \forall \ p\in[\underline p, \overline p]
\end{equation}where $ R(p',p;\beta)=\sum_{s=1}^\infty (-\beta)^{s-1} K_s(p',p;\beta)$, in which  $ K_s(p',p;\beta)=\int_0^1  K_{s-1}(p',\tilde p;\beta)\cdot \Pi(\tilde p, p;\beta)d\tilde p$ and $ K_1(p',p;\beta)=\Pi(p',p;\beta) $. 
\end{lemma}


\noindent
The solution \eqref{eq8} is proportional to $\theta$, which is due to the linearity of FIE system.  Therefore, \eqref{eq8} can also be represented by a sequence of ``basis'' solutions.  To see this, let $ z_\ell(p)$ be the $\ell$--th argument of $z(p)$. For $\ell =1,\cdots,k_\theta$, let $b^*_\ell(\cdot)$ be the (unique) solution to the following equation
\begin{equation}
\label{eq9}
b_\ell(p)+\beta\int_{\underline p}^{\overline p}\int_{\underline p}^{p'}b_\ell(\tau)d\tau \cdot \pi(p',p;\beta) dp' 
=z_\ell(p).
\end{equation} By a similar argument to \Cref{lemma4},  we have 
\[
b^*_\ell(p)=z_\ell(p)-\beta\int_{\underline p}^{\overline p}  R(p',p;\beta)\cdot z_\ell(p') dp',\ \ \ \forall \ p\in[\underline p, \overline p]
\] as the unique solution to \eqref{eq9}. Let $\mathcal B(\cdot)\equiv(b^*_1(\cdot),\cdots,b^*_{k_\theta}(\cdot))^\intercal$ be the sequence of solutions supported on $[\underline p, \overline p]$. Thus, the solution in eq. \eqref{eq8} can be written as 
\begin{equation}\label{qb}
Q(p)=\mathcal B(p)^\intercal\cdot \theta, \ \ \forall \ p\in[\underline p, \overline p]
\end{equation}

By \Cref{lemma1,lemma2,lemma3,lemma4}, we obtain a single--index representation of the semiparametric dynamic decision model, which is stated in the following theorem.
\begin{theorem}
\label{th1}
Suppose \cref{as1,as_regular,as2,as3,as4} hold. Then, the agent's dynamic decision can be represented by a static single--index model:
\[
\Pr(Y=1|X)=F_{\eta}\big(m(X)^\intercal\cdot\theta\big)
\]where 
\[
m(X)=\phi(X)-\sum_{s=1}^\infty \beta^{s} \left\{\mathbb E \Big[\int_{\underline p}^{p(X^{[s]})}\mathcal B(\tau)d\tau \big|X,Y=1\Big]-\mathbb E \Big[\int_{\underline p}^{p(X^{[s]})}\mathcal B(\tau)d\tau \big|X,Y=0\Big]\right\}.
\]
\end{theorem}
\noindent Note that $\Pr(Y=1|X)=F_{\eta}\big(Q(p(X))\big)$. Then, \Cref{th1} obtains by combining eqs. \eqref{eq6}. and \eqref{qb}. Given the identification of $\mathcal B(\cdot)$ on the support $[\underline p, \overline p]$,  $m(\cdot)$ is then constructively identified on $\mathscr S_X$. Therefore, the identification of $\theta$ simply follows the single-index model literature, see e.g. \cite{manski1975maximum,manski1985semiparametric}.

It is worthing noting that  any constant term in $W_d$ remains a constant
in the transformed linear--index $m(X)$. In other words, suppose, w.l.o.g., $W_{11}=1$, then the corresponding argument in $m(X)$ also equals 1. To see this, note that the first argument in $\phi(X)$ is given by
\[
 1+ \sum_{s=1}^\infty\beta^s\left\{\mathbb E \big[p(X^{[s]})|X,Y=1\big]-\mathbb E \big[p(X^{[s]})|X,Y=0\big]\right\},
\]which thereafter implies 
\[
z_1(p)= 1+ \sum_{s=1}^\infty\beta^s\left\{\mathbb E \big[p(X^{[s]})|p(X)=p,Y=1\big]-\mathbb E \big[p(X^{[s]})|p(X)=p,Y=0\big]\right\}.
\]
Using \eqref{eq9}, one can verify that the solution is: $b^*_1(\cdot)=1$. Then, we plug this solution into the the first element of $m(X)$, which gives us  $m_1(X)=1$. By a similar argument, a constant in $W_{0}$ implies the corresponding term in $m(X)$ equals $-1$. 

By a similar argument as in the static binary response model literature,
the index parameter $\theta$ is identified up to location and scale in the
semiparametric setting. For notational simplicity, hereafter we assume the
state vector $X$ does not include a constant term in the semiparametric
setting.\footnote{In our semiparametric setting, any constant term in the
  utility function will be absorbed by the error term since the
  distribution of the latter is left unspecified.} Moreover, we will
introduce a scale normalization on $\theta$ which is also standard in the
literature.

\begin{assumption}
\label{as5}
We denote the first argument of $m(X)$ by $m_1(X)$ and the rest by $m_{-1}(X)$. Moreover, let $m_{1}(X)$ be continuously distributed on an interval
given $m_{-1}(X)$ which is a vector of either discrete and/or continuous random variables.  Let $f_{m_1(X)|m_{-1}(X)}$ be the conditional pdf. Moreover, the matrix $\mathbb E [m(X)m(X)^\intercal]$ is invertible.  
\end{assumption}

\noindent
In \Cref{as5}, the first half condition requires at least one argument in $X_1$ to be continuously distributed conditional on others; this rules out cases where, e.g. all the state variables are functions of a single variable $X_1$ (as in \citep{rust1987optimal}, where mileage and mileage-squared enter as state variables). The second half of \Cref{as5} is a testable rank condition. \Cref{as5} is a strong assumption, but almost indispensable in the semiparametric single index model literature; See \cite{horowitz2009semiparametric}. 
\begin{assumption}
\label{as6}
Let $\|\theta\|=1$.
\end{assumption}
\noindent
\Cref{as6} is a scale normalization, which has also been used in PSS. Note that we implicitly normalize our location term by 0, since neither $W_0$ nor $W_1$  contains a constant term. 


\begin{theorem}
\label{theorem1}
Suppose \cref{as1,as_regular,as2,as3,as4,as5,as6} hold.  Then, the structural parameter $\theta$ is point identified.
\end{theorem}

\section{Semiparametric Estimation}
In this section, we  describe and motivate the semiparametric estimation of our structural model. For expositional simplicity,  we assume all variables in $X$ are continuously distributed. A mixture of continuous and discrete regressors can be accommodated at the expense of notation.  Let $\{(Y_t, X^\intercal_t)^\intercal: t=1,\cdots, T\}$ be  our  sample of the Markov decision process.  Our estimation procedure simply follows the identification strategy, which takes multiple steps. Throughout, we use  $K$ and $h$ to denote a  Parzen--Rosenblatt kernel and a bandwidth, respectively.

 First, we nonparametrically estimate the choice probabilities $p(\cdot)$ and the generated regressor $\phi(\cdot)$. In particular, let
\[
\hat p(X_s)= \frac{\sum_{t=1}^T Y_t\times K_p\Big(\frac{X_t-X_s}{h_p}\Big)}{\sum_{t=1}^T  K_p\Big(\frac{X_t-X_s}{h_p}\Big)}, \ \ \forall s=1,\cdots, T.
\]As standard, we choose an optimal bandwidth, i.e.,   $h_p= 1.06\times \hat\sigma(X)\times T^{-\frac{1}{2\iota +k}}$, where $\hat\sigma(X)$ is the sample standard deviation of $X_t$ and $\iota $ ($\iota\geq 2$) is the order of the kernel function $K_p$. For example, if we choose $K_p$ to be the pdf of the standard normal distribution, then $\iota=2$. In addition, the support $[\underline p, \overline p]$ of $p(X)$ can be estimated by $[\min_{1\leq s\leq T} \hat p(X_s), \max_{1\leq s\leq T}\hat p(X_s)]$. 

Moreover, recall that the transformed state variables $W_d(X)$ ($d=0,1$) are known. Then, for $s=1,\cdots,S_T$, where $S_T=T-\ell_T$ for some integer $\ell_T$ satisfying $\ell_T\rightarrow+\infty$ and $S_T\rightarrow+\infty$ as $T\rightarrow +\infty$, let 
$\delta_{dt}= \sum_{s=1}^{\ell_T}\beta^s\cdot W_{d}(X_{t+s}) Y^d_{t+s}(1-Y_{t+s})^{1-d}$. For $s=1,\cdots,T$, let further 
\[
\hat \phi_d(X_s)= (-1)^{d+1} W_{d}(X_{s})
+\frac{\sum_{t=1}^{S_T} \delta_{dt}\cdot K_\phi\Big(\frac{X_t-X_s}{h_\phi}\Big)\mathbb 1 (Y_t=1)}{\sum_{t=1}^{S_T} K_\phi\Big(\frac{X_t-X_s}{h_\phi}\Big)\mathbb 1 (Y_t=1)}-\frac{\sum_{t=1}^{S_T} \delta_{dt}\cdot K_\phi\Big(\frac{X_t-X_s}{h_\phi}\Big)\mathbb 1 (Y_t=0)}{\sum_{t=1}^{S_T} K_\phi\Big(\frac{X_t-X_s}{h_\phi}\Big)\mathbb 1 (Y_t=0)}.
\]Similarly, we can choose  $h_\phi$ in an optimal way.  In above expression, the summation includes only  the first $S_T$ observations. This is because $\delta_{dt}$ is not well defined for all $t>S_T$. In practice, we choose $\ell_T$ in a way such that 
$\delta_{dt}- \sum_{s=1}^{+\infty}\beta^sW_{d}(X_{t+s}) Y^d_{t+s}(1-Y_{t+s})^{1-d}$ is negligible relative to the sampling error, which is feasible because the former converges to zero at an exponential rate. 

In the second stage, we estimate $z(\cdot)$ and $\mathcal B(\cdot)$ on the support $[\underline p, \overline p]$. First, let  
\[
\hat z(p)=\frac{\sum_{t=1}^{T} \hat \phi(X_{t}) \cdot K_z\Big(\frac{\hat p(X_t)-p}{h_z}\Big)}{\sum_{t=1}^{T}K_z\Big(\frac{\hat p(X_t)-p}{h_z}\Big)},\ \ \forall \  p\in [\min_{1\leq s\leq T} \hat p(X_s), \max_{1\leq s\leq T}\hat p(X_s)].
\] According to \cite{vuong_perrigne_guerre}, $h_z$ is chosen in a suboptimal way, i.e., $h_z=1.06\times \hat\sigma(p(X)) \times T^{-\frac{1}{2\iota+3}}$. 

To estimate $b^*_\ell(\cdot)$ on the support $[\underline p, \overline p]$, we note that eq. \eqref{eq9} can be rewritten as
\begin{multline*}
b_\ell(p)+\sum_{s=1}^\infty\beta^s\cdot  \mathbb E \Big[\int_{\underline p}^{p(X^{[s]})}b_\ell(\tau)d\tau\big|p(X)=p,Y=1\Big]\\
 -\sum_{s=1}^\infty\beta^s\cdot  \mathbb E \Big[\int_{\underline p}^{p(X^{[s]})}b_\ell(\tau)d\tau\big|p(X)=p,Y=0\Big] 
=z_\ell(p).
\end{multline*}
This suggests an estimator $\hat b^*_\ell(\cdot)$ that solves
\[
\hat b^*_\ell(p)+\frac{\sum_{t=1}^{S_T}\xi_{t}(\hat b^*_\ell) \times K_\xi\Big(\frac{\hat p(X_t)-p}{h_\xi}\Big)\times Y_t}{\sum_{t=1}^{S_T}K_\xi\Big(\frac{\hat p(X_t)-p}{h_\xi}\Big)\times Y_t}-\frac{\sum_{t=1}^{S_T}\xi_{t}(\hat b^*_\ell) \times K_\xi\Big(\frac{\hat p(X_t)-p}{h_\xi}\Big)\times (1-Y_t)}{\sum_{t=1}^{S_T}K_\xi\Big(\frac{\hat p(X_t)-p}{h_\xi}\Big)\times (1-Y_t)}
=\hat z_\ell(p),
\]
where $\xi_{t}(b_\ell)=\sum_{s=1}^{\ell_T}\beta^s \int_{\underline p}^{\hat p(X_{t+s})} b_\ell(\tau)d\tau$ for which the integration can be computed by the numerical integration. Similarly, $h_z=1.06\times \hat\sigma(p(X)) \times T^{-\frac{1}{2\iota+3}}$  is chosen sub-optimally.  A numerical solution of $\hat b^*_\ell$ can obtain using the iteration method: Let $\hat b^{[0]}_\ell=\hat z_\ell^\intercal(p)$. Then we set
\[
\hat b_\ell^{[1]}(p)
=\hat z_\ell^\intercal(p)-\left\{\frac{\sum_{t=1}^{S_T}\xi_{t}(\hat b^{[0]}_\ell) \times K_\xi\Big(\frac{\hat p(X_t)-p}{h_\xi}\Big)\times Y_t}{\sum_{t=1}^{S_T}K_\xi\Big(\frac{\hat p(X_t)-p}{h_\xi}\Big)\times Y_t}-\frac{\sum_{t=1}^{S_T}\xi_{t}(\hat b^{[0]}_\ell) \times K_\xi\Big(\frac{\hat p(X_t)-p}{h_\xi}\Big)\times (1-Y_t)}{\sum_{t=1}^{S_T}K_\xi\Big(\frac{\hat p(X_t)-p}{h_\xi}\Big)\times (1-Y_t)}\right\}.
\]Repeat such an iteration until it converges. Then we obtain $\hat b^*_\ell(\cdot)=\hat b_\ell^{[\infty]}(\cdot)$ on $[\hat {\underline p},\ \hat {\overline p}]$.

Next, we obtain the single--index variables $m(X_s)$ by: for $\ell =1,\cdots, k_\theta$,
\[
\hat m_\ell(X_s)=\hat \phi_\ell(X_s) - \left\{\frac{\sum_{t=1}^{S_T}\xi_{t}(\hat b^*_\ell) \times K_m\Big(\frac{X_t-X_s}{h_m}\Big)\times Y_t}{\sum_{t=1}^{S_T}K_m\Big(\frac{X_t-X_s}{h_m}\Big)\times Y_t}-\frac{\sum_{t=1}^{S_T}\xi_{t}(\hat b^*_\ell) \times K_m\Big(\frac{X_t-X_s}{h_m}\Big)\times (1-Y_t)}{\sum_{t=1}^{S_T}K_m\Big(\frac{X_t-X_s}{h_m}\Big)\times (1-Y_t)}\right\}. 
\] In particular, $h_m= 1.06\times \hat\sigma(X)\times T^{-\frac{1}{2\iota +k}}$ is chosen optimally.

Finally, we apply PSS to estimate $\theta$ (up to scale).\footnote{One could also use alternative methods e.g. \cite{klein1993efficient} and \cite{ichimura1993semiparametric} to estimate $\theta$.}  Specifically, we define
\begin{equation}\label{pssestimator}
\hat \theta =-\frac{2}{T(T-1)}\sum_{t=1}^T \sum_{s\neq t}\left[\frac{1}{h^{k_\theta+1}_\theta}\times \nabla K_\theta\left(\frac{\hat m(X_t)-\hat m(X_s)}{h_\theta}\right)\times Y_s\right].
\end{equation} Following the standard kernel regression literature, we can show $\hat \theta$ is consistent given that $\sup_{x\in\mathscr S_X}|\hat m(x)-m(x)| = o_p(h_\theta)$, $h_\theta\rightarrow 0$ and $Th^{k_\theta+1}\rightarrow \infty$ as $T\rightarrow \infty$. 

Similar to PSS, it is of particular interest to establish $\sqrt T$--consistency of $\hat \theta$. The argument follows closely that in PSS.  In particular, we need choose a high order kernel $K_\theta$  and an under--smoothed bandwidth $h_\theta$. However,  it is more delicate in our setting because of the generated regressor  $\hat{m}(X)$ contained in the kernel function of our estimator \eqref{pssestimator}. Due to the first--stage estimation error, we must make the additional assumptions on the convergence of $\hat{m}(X)$ to $m(X)$:
\begin{assumption}\label{as9}
$h_\theta=T^{-\frac{1}{\gamma}}$ where
$k_\theta+2<\gamma<k_\theta+3+\mathbbm{1}(k_\theta \text{ is even})$.
\end{assumption}
\begin{assumption}\label{as10}
The support of the kernel function $K_\theta$ is a convex subset of $\mathbb R^{k_\theta}$ with nonempty
interior, with the origin as an interior point. $K_\theta$ is a bounded differentiable function that obeys:  $\int K_\theta(u)du=1$, $K_\theta(u)=0$ for all u belongs to the boundary of its support, $K_\theta(u)=K_\theta(-u)$ and 
\begin{align*}
&\int u^{\ell_1}_1\cdots u^{\ell_{\rho'}}_{k_\theta}K_\theta(u)du=0, \ \ \ \text{for }\  \ell_1+\cdots+\ell_{\rho'}<\frac{k_\theta+3+\mathbb 1 (k_\theta \text{ is even })}{2}, \ \text{and}\\
&\int u^{\ell_1}_1\cdots u^{\ell_{\rho'}}_{k_\theta}K_\theta(u)du\neq 0, \ \ \ \text{for }\  \ell_1+\cdots+\ell_{\rho'}=\frac{k_\theta+3+\mathbb 1 (k_\theta \text{ is even })}{2}.
\end{align*}where $u_\ell$ is the $\ell$--th argument of $u$.
\end{assumption}

%
%


\begin{assumption}\label{as8}
(i) $\mathbb E  \|\hat m(X)-m(X)\|^2=o(T^{-\frac{1}{2}}h^{3}_\theta)$;\\
(ii) $\mathbb E  \|\mathbb E[\hat m(X)|X]-m(X)\|\ =o(T^{-\frac{1}{2}}h^{2}_\theta)$;\\
(iii) $\hat m(X_t) - \hat m_{t,-s} =o_p(T^{-\frac{1}{2}}h^{2}_\theta)$, where $\hat m_{t,-s}$ is the nonparametric estimator $\hat m(X_t)$, except for leaving the $s$--th observation out of the sample in its construction. 
\end{assumption}
\noindent
\Cref{as9,as10} are introduced by PSS for the choice of  bandwidth and kernel, respectively, to control the bias term in the estimation of $\theta$.\footnote{Note that we implicitly assume that Assumptions 1 -- 3 in PSS hold, which impose  smoothness conditions on $f_{m(X)}$ and $\Pr(Y_t=1|m(X_t)=m)$ as well as other regularity conditions.} The restriction on the bandwidth \Cref{as9} implies that $h_{\theta}$ is not an optimal bandwidth sequence (rather it is undersmoothed) such that the bias of estimating $\theta$ goes to zero faster than $\sqrt T$. Moreover, \Cref{as8} encompasses high--level conditions that could be further established under primitive conditions.  In particular, \Cref{as8}(i)  requires $\hat m(\cdot)$ converge to $m(\cdot)$ faster than $T^{-\frac{1}{4}}$. Similar conditions are also made in e.g. \cite{ai2003efficient} for the regular convergence of finite--dimensional parameters  in  semiparametric models. Moreover, by \Cref{as8}(ii), the bias term in the estimation of $m$ uniformly converges to zero faster than $T^{-\frac{1}{2}}$. Hence, we need to use  higher order kernel  in the estimation of $m(\cdot)$. \Cref{as8}(iii) is not essential, which could be dropped if we exclude both $t$-th and $s$--th observations in the argument $\hat m(X_t)-\hat m(X_s)$  of the kernel function in \eqref{pssestimator}.

Given these assumptions, we can show the following result (the proof is in the appendix):
\begin{theorem}
\label{th2}
Suppose \cref{as8,as9,as10} hold.   Then, for some scalar $\lambda> 0$ specified below, $\sqrt T (\hat{\theta}-\lambda\cdot \theta)$ has a limiting
multivariate normal distribution  defined in  \citet[Theorem 3.1]{powell1989semiparametric}:
\[
\sqrt T (\hat{\theta}-\lambda\cdot \theta)\overset{d}{\rightarrow } N(0,\Sigma)
\]where $\Sigma\equiv 4\ \mathbb E (\zeta\cdot \zeta^\intercal)-4\lambda^2\times \theta \cdot \theta^\intercal$, $\zeta=f_{m}(m(X))\cdot f_\eta(\eta^*(X))\cdot \theta-\big[Y-F_\eta(\eta^*(X))\big]\cdot f'_{m}(m(X))$ and  $\lambda=\mathbb E \big[f_{m}\big(m(X)\big)\times f_{\eta}(m(X)^\intercal\cdot\theta)\big]$. 
\end{theorem}
\noindent
 In the above theorem, recall $\Pr(Y=1|X)=F_\eta(\eta^*(X))$ and $\eta^*(X)=m(X)^\intercal\cdot \theta$ by \Cref{th1}.  
Our estimator $\hat\theta$ (as defined in Eq. (\ref{pssestimator})) has not imposed the scale restriction in \Cref{as6}; thus $\lambda\in\mathbb R$ in the above theorem denotes the probability limit of $\|\hat{\theta}\|$; i.e., $\|\hat \theta\|=\lambda+O_p(T^{-1/2})$.  
Therefore, by rescaling our estimator $\hat\theta$ as $\hat\theta^*={\hat \theta}/{\lambda}$, we obtain that
\[
\sqrt T (\hat{\theta}^*- \theta)\overset{d}{\rightarrow } N(0,\Sigma/\lambda^2).
\]

Given $\hat\theta^*$, a  nonparametric estimator of $Q(\cdot)$ directly follows from Eq. \eqref{eq8}. Namely, let 
\[
\hat Q(p)= \hat z^\intercal(p)\times \hat \theta^*, \ \ \forall \  p\in [\min_{1\leq s\leq T} \hat p(X_s), \max_{1\leq s\leq T}\hat p(X_s)].
\]Because of the $\sqrt T$--consistency of $\hat \theta^*$, the estimator $\hat Q_p(p)$ is asymptotically equivalent to $ \hat z^\intercal(p)\times \hat\theta^*$, which converges at a nonparametric rate. 

\subsection{Monte Carlo}
The focus of our Monte Carlo is on the semiparametric estimation. In our  experiments, let  $u_t(0,X_t,\epsilon_t)=\theta_0+\epsilon_{0t}$ and $u_t(1,X_t,\epsilon_t)=X_{1t}\theta_1+X_{2t}\theta_2+\epsilon_{1t}$, where $X_{1t},X_{2t}$ are random variables and $\theta_0,\theta_1,\theta_2\in\mathbb R$.  Moreover, we set the conditional distribution of $X_{t+1}$ given $X_t$ and $Y_t$ as follows: for $k=1,2$
$$
X_{k,t+1}=\left\{\begin{array}{cc}X_{kt}+\nu_{kt}, & \text{if }Y_t=0 \\\nu_{kt} & \text{if }Y_t=1\end{array}\right.
$$
where $\nu_{kt}$ conforms to $\ln \mathcal N(0,1)$ and $\nu_{1t}\bot \nu_{2t}$. Moreover, let $\epsilon_{dt}$ be i.i.d. across $d=0,1$ and $t$, and conform to an extreme value
distribution with the density function $f(e)=\exp(-e) \exp[-\exp(-e)]$.  We set $\beta=0.9$ and the parameter value as follows: $\theta_0=-5$, $\theta_1=-1$ and $\theta_2=-2$. 

Because we cannot estimate the constant $\theta_0$ in the semiparametric framework, then we treat $\theta_0$ as a nuisance parameter. Let $\theta=(\theta_1,\theta_2)^\intercal$. As a matter of fact,  $\theta$ is only  identified up to scale in the semiparametric setting. To compare the performance of the semiparametric estimators, we assume the scale of $\theta$ is known, i.e., $\|\theta\|=\sqrt {5}$, rather than imposing a different normalization, as \cref{as6}.  We present in Table \ref{table:mc1} the bias and standard deviation of the semiparametric estimator.

\begin{table}[t]

\begin{center}
\begin{tabular}{ c c c  c c c  }
\toprule

\noalign{\smallskip}
Sample Obs. & Parameter  & True Value            & Estimate  & Std. Dev.	& Bias    \\
\midrule

\multirow{2}{*}{1000} & $\theta_1$ & $-1$        & -1.0182   &    0.3636	       & 0.0182 \\
                                & $\theta_2$ & $-2$        & -1.9457   &    0.2158         & -0.0543	\\
                                
                                &&&&&\\

\multirow{2}{*}{2000} & $\theta_1$ & $-1$        & -1.0163   &   0.2913 	& 0.0163 \\
                                  & $\theta_2$ & $-2$        & -1.9618   &   0.1854        & -0.0382	\\
                                  
                                &&&&&\\

\multirow{2}{*}{4000} & $\theta_1$ & $-1$        & -0.9985   &   0.2344 	& -0.0015 \\
                                  & $\theta_2$ & $-2$        & -1.9836   &   0.1176         & -0.0164	\\

\bottomrule
\end{tabular}

\caption{Monte Carlo Results}
\label{table:mc1}
\floatfoot{This table presents Monte Carlo results for different sample sizes. For each sample size, reported estimates, standard deviations and bias are computed as the mean across 150 simulation draws.}

\end{center}
\end{table}

\section{Empirical application: Dynamic labor supply for NYC taxi drivers}

Several recent papers seek to estimate labor supply elasticities in markets where labor supply is continuously adjustable. Most of these papers are centered around the market for taxi rides, because taxi drivers choose their own hours. This analysis aims to characterize the dynamic labor supply of taxi drivers. We first pose and estimate a model of taxi driver's labor supply as a dynamic discrete choice over quitting for the day. Our model highlights the tradeoffs between working longer to earn extra income versus incurring increasing costs of effort. Estimating the structural model allows us to both analyze the labor-leisure tradeoff in a richer way compared to previous studies, and also to showcase the features of our semiparametric estimation procedure.

This work builds on a literature on labor supply in the taxi industry. \citet*{farber_2005, farber_2008, farber_2014} analyze the labor supply of New York City taxi drivers. To contrast with previous literature which found small labor supply elasticities among workers, Farber points out that most workers are not free to set their own hours. Taxi drivers are an ideal subject to study these elasticities precisely because they are free to set their own hours. He finds that drivers' labor supply is consistent with the predictions of standard neoclassical models of inter-temporal labor supply; increases in wage rates correspond to drivers working longer hours. A puzzle remained however, that Farber's study did not corroborate earlier work on Taxi drivers, notably \citet*{camerer1997labor}, which found strong negative wage elasticities. Negative elasticities could reflect the presence of income-targeting by drivers: for example, a labor supply policy of the form ``I will work today until I earn \$200." 

Using the same Taxi data, Farber's later work responds to the discrepancy of Camerer et. al and the question of income-targeting by estimating a model that integrates reference-dependent utility. Reference-dependence in utility is the notion that agents' utility is not only a function of income but also reference-points or targets, where the marginal utility of income increases more quickly before the target is met than after it is met. Originally, \citet*{farber_2008} finds mixed evidence for the existence of reference-dependence, but \citet*{farber_2014} uses more comprehensive data and finds strong evidence that labor supply behavior is driven by the standard neoclassical prediction of upward sloping supply curves, as opposed to income-targeting and its associated negative elasticities. 

In this analysis, we ask how taxi drivers choose to stop work for the day by estimating a dynamic structural model in which drivers' stopping decision is a function of both cumulative earned income and cumulative time spent working. Our model is based on the taxi labor supply model of \citet*{frechette2015} [FLS], in which taxi drivers solve a dynamic competitive game by choosing the optimal starting times and length of time to stay on a shift.\footnote{The more taxis that are working, the less revenue is earned as a result of lower probabilities of finding a passenger.} As with FLS, our taxi drivers will decide how long to work by weighing the utility of earning revenue against the disutility of working longer. FLS utilizes the MPEC method to solve a dynamic entry game in an equilibrium framework, allowing the market to equilibrate via the waiting times experienced by passengers and taxis. While we do not consider these general equilibrium forces, we take advantage of our computationally light, semi-parametric estimation method to test for a variety of taxi driver stopping behaviors posited by previous authors. Thus our approach is partial equilibrium, as agents in our model take the waiting times and arrival of customers as given rather than endogenously determined in a dynamic equilibrium (as in FLS).

Both models can also be viewed as a stopping rule framework akin to the classic paper by \citet*{rust1987optimal}. Rust models the decision to replace bus engines which weighs routine maintenance costs against a higher probability of catastrophic engine failure. In this setting, after each trip given, a taxi driver must weigh the opportunity for additional fares against a rising cost of fatigue in each day.\footnote{In other words, drivers experience increasingly large marginal utility of leisure as the available hours of leisure drops.} We use our estimation method to recover parameters defining driver's cost functions.

\subsection{Behavioral model}

\subsubsection{Period revenues and costs}

Taxi drivers assumed to have costs of effort that are increasing in hours worked each day.  After each ride given, drivers face a discrete decision, to continue searching for passengers or quit for the day. In this sense, their labor supply decision boils down to a comparison between the expected profit of searching for a unit of time versus the disutility of driving for that much more time. The period payoff function for driver $i$ depends on the decision to quit ($y_{it}=1$) or keep working ($y_{it}=0$) takes the following form:

\begin{equation}
u_i(s_{it}, h_{it},y_{it}; \theta, X_t)=\begin{cases}

\begin{array}{c}


\theta_u \cdot s_{it}  + \varepsilon_i(1) \\

\theta_{c,01} \cdot h_{it} +  \theta_{c,02} \cdot h_{it}^2 +\varepsilon_i(0)

\end{array} & \begin{array}{c}

\mbox{if }y_{it}=1\\
\mbox{if }y_{it}=0
\end{array}\end{cases}.
\end{equation}

This dynamic labor supply model is an optimal stopping model, in which the taxi driver's dynamic problem ends once he decides to end his current shift.   The terminal utility from ending the shift is given in the upper ``prong'' of the utility specification above.  In this terminal utility, the term $\theta_u \log(s_{it})$ captures the utility from earnings enjoyed by the taxi driver after ending his shift, and $\theta_{c,1} \cdot h_{it}$ likewise captures the post-shift utility depending on the cumulative hours worked.   When a driver continues to drive ($y_{it}=0$), our specification assumes that he receives (dis-)utility from doing so, which depends on the cumulative hours worked so far in this shift.  This $\theta_{c,0}\cdot h_{it}$ measures the cost of the effort exerted by the working driver, which may change as the cumulative hours $h_{it}$ increases.

%
%

\subsection{Data}

In 2009, The Taxi and Limousine Commission of New York City (TLC) initiated the Taxi Passenger Enhancement Project, which mandated the use of upgraded metering and information technology in all New York medallion cabs. The technology includes the automated data collection of taxi trip and fare information. We use TLC trip data on all New York City medallion cab rides given in February, 2012.  The sample analyzed here consists of 44,090 observations, or about 0.3\% of the data. Each row in the data is information related to a single cab ride.  Data include driver and medallion identifiers, the exact time and date of pickup and drop-offs, trip distance, and trip time for approximately 44,090 individual taxi rides. Table \ref{table:sumstat1} provides summary statistics.

\begin{table}[t]

\begin{center}
\begin{tabular}{ l c  c c c c c  }
\toprule

\noalign{\smallskip}
Trips/Shifts & Variable                          & Obs.	              & 10\%ile	& Mean    & 90\%ile	& S.D. \\
\midrule

\multirow{2}{*}{\parbox{2.5cm}{Trip Statistics}} & Trip Revenue	(\$)         & 44,090   &	 5.52 	& 12.36	& 21.74	& 9.64 \\
& Trip Time (min.)	 & 44,090   &      3.67         & 11.50	& 22.0	& 8.33 \\

\midrule
\multirow{2}{*}{\parbox{2.5cm}{Shift Statistics}} & Shift Revenue (\$)            	& 1857       & 163.58	& 293.37	& 417.09	& 138.61 \\
& Shift Time  (min.)             	& 1857     & 300.37	& 529.53	& 689.20	& 287.34	\\

\bottomrule
\end{tabular}

\caption{Taxi Trip and Fare Summary Statistics}
\label{table:sumstat1}
\floatfoot{Taxi trip and fare data come from New York Taxi and Limousine Commission (TLC) and refer to February 2012 data. The first set of statistics relates to individual taxi trips. The second set of statistics relate to cumulative earnings and time spent in individual driver shifts.}

\end{center}
\end{table}


\subsection{Results}

The estimation results are presented in Table \ref{tab:estimates1}. For estimation, we scaled the cumulative time variable to be in units of five-minutes. We find that the terminal utility upon ending a shift grows with log earnings, which is weighed against a negative effect of cumulative hours worked, which accumulates in each period of continued work. These estimates highlight a classic upward-sloping labor supply curve, which is manifested here as a stopping rule in which total hours worked is increasing in income. It is important to note that the relatively small coefficient on hours worked is to be expected, since this utility accrues in every period that a driver continues working, while the utility benefit of earned income is only received once, when the driver stops working for the day. Indeed, we cannot infer the ``relative importance" of hours vs. income directly from the estimates, since the utility of each accrues differently. 



\begin{figure}[!t]
	\caption{Estimated Quantile Function
	\label{fig:Q_estimate1}} 
		\includegraphics[scale=.65]{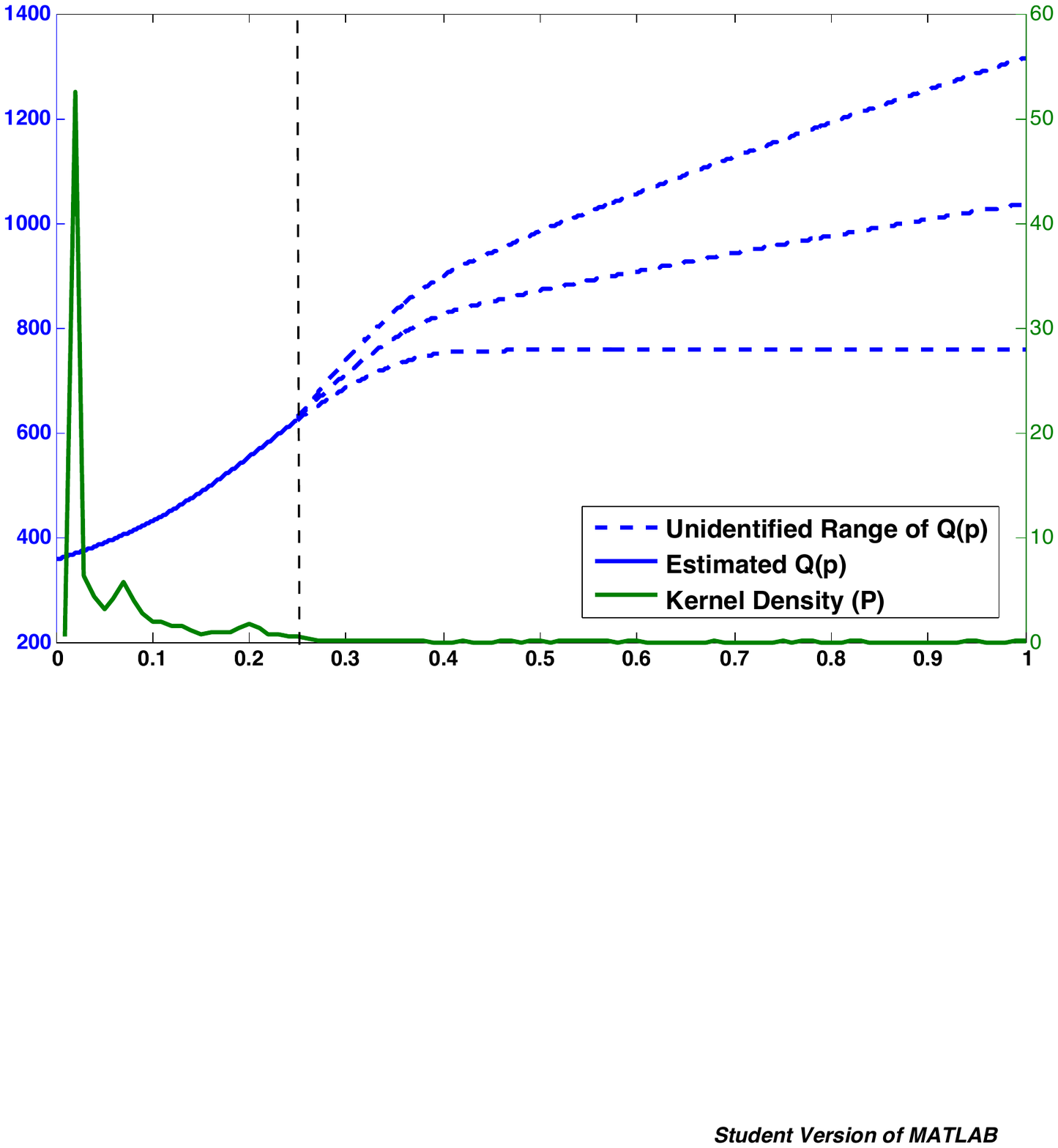} 		
\end{figure}

Given these parameter estimates, in Figure \ref{fig:Q_estimate1} we graph the implied quantile function for the difference in utility shocks $\eta\equiv \epsilon_1-\epsilon_0$. The density of $\hat{p}$ is plotted as well, which highlights a range over which choice probabilities are actually observed. Outside of this range, we are unable to identify the corresponding quantile function, which is depicted here as flat. Using the density of $\hat{p}$ as a guide, we can recover the quantile function for the range of percentiles approximated by [0.05, 0.25]. A thin vertical dotted line depicts this range. Note that the shocks take (even very large) positive values with about 95\% probability.   This may imply that there is a large fixed positive component to the terminal utility from quitting. 

\begin{figure}[!t]
	\caption{Estimated Quantile Function (Trimmed)
	\label{fig:Q_estimate2}} 
		\includegraphics[scale=.65]{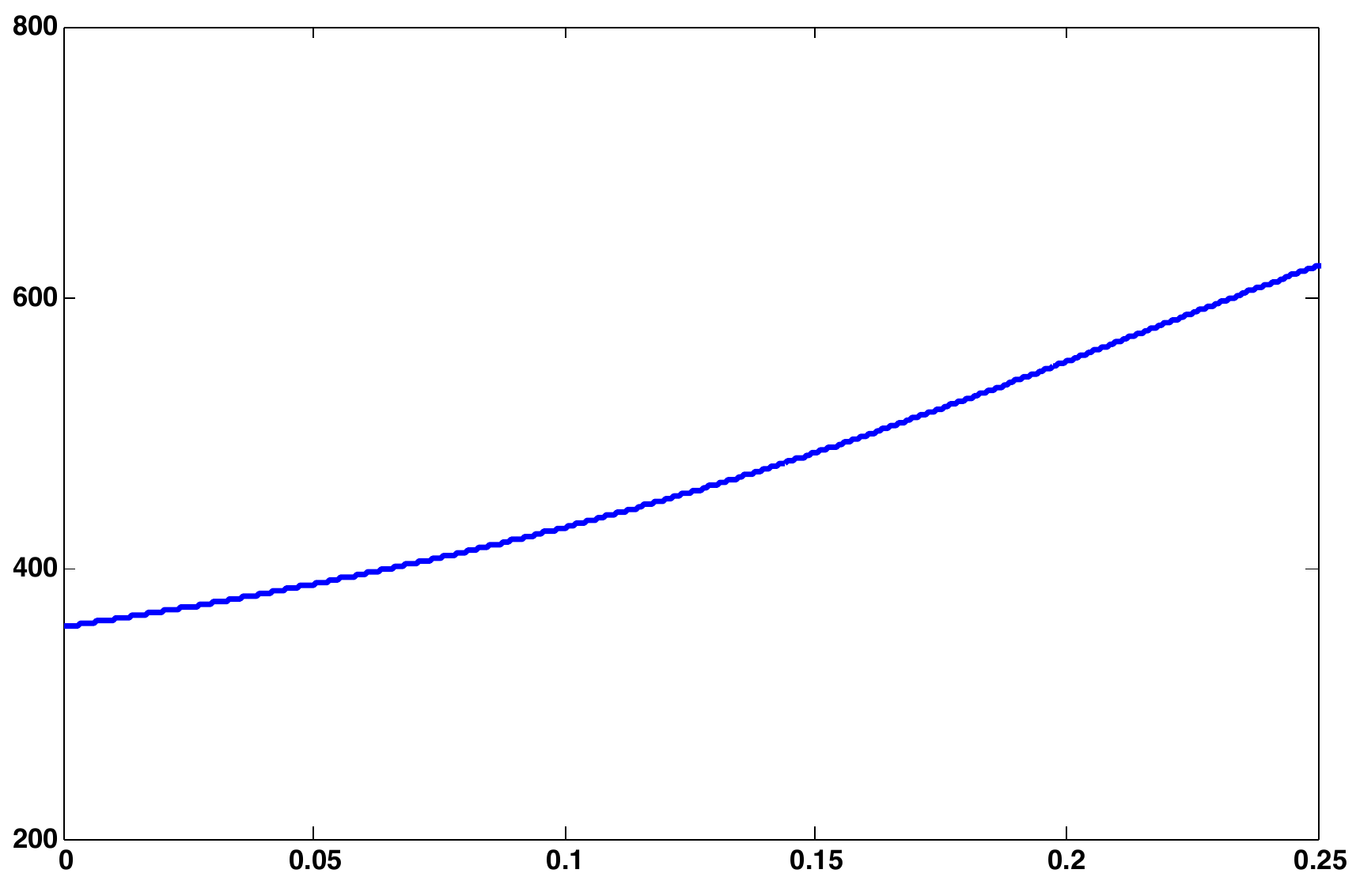} 		
\end{figure}

This feature that, as shown in Figure \ref{fig:Q_estimate2}, our approach only yields an incomplete estimate of this distribution, may be problematic for evaluating some counterfactual policies.  For certain counterfactuals, knowledge of the entire distribution of the utility shocks is required, since this distribution feed agents' beliefs about the future.  In ongoing work, we are exploring ways for extrapolating this distribution beyond the range identified by our approach.

%
%
%

\begin{table}[ht]
\begin{center}
\caption{Parameter Estimates \label{tab:estimates1}}

\begin{tabular}{c  l  c c  }
\toprule

 \textbf{Parameter}        & \textbf{Description}   &  \textbf{Estimate} & \textbf{Std. Error} \\ 
\midrule
$\theta_u$                      &    Earnings (upon quitting)                 &  $0.9907$       &   0.0118      \\ 
 $\theta_{c,01}$              &   Cumul. hours (while working)                & $-0.1359$      &   0.0759 \\ 
 $\theta_{c,02}$              &   Cumul. hours squared (while working)  &   $-0.0004$    &   0.0002       \\ 

\bottomrule
\end{tabular}
\floatfoot{Note: Standard errors are computed by estimating the model 200 times and reporting the standard deviation of estimates.}
\end{center}
\end{table}

\begin{figure}[!t]
	\caption{Estimated Choice-specific Value Function Differences: $V_1(X)-V_0(X)$
	\label{fig:mtheta}} 
		\includegraphics[scale=.85]{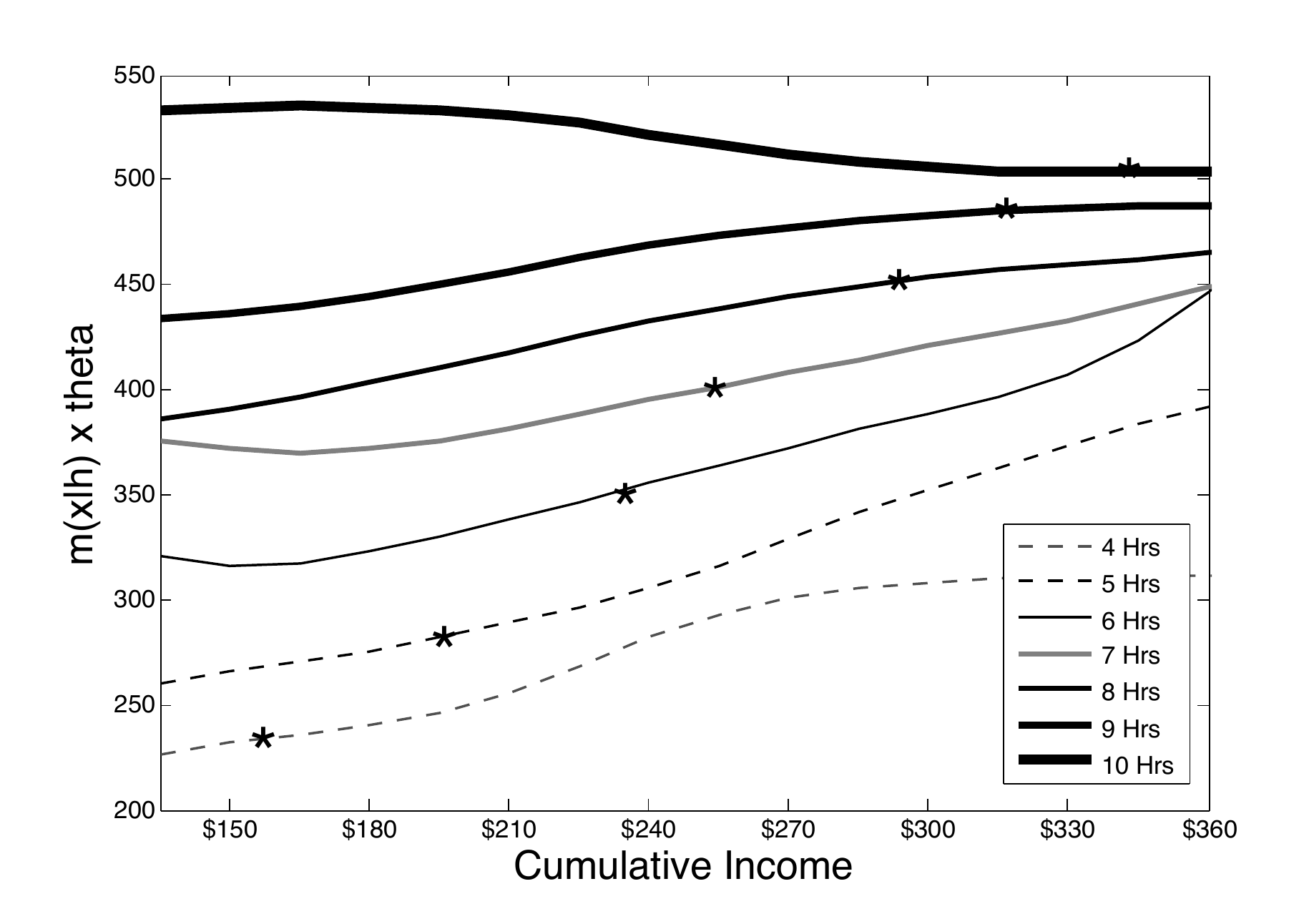} 		
		\floatfoot{We report the estimates of the $m(X)'\theta$ function, as in Theorem~\ref{th1}, for $X=(hours\ worked, income)$.   For fixed values of hours worked, we graph $m(X)'\theta$ as a function of income.   Stars (*) mark average income earned by drivers for a given hours worked, as observed in the raw data.}
\end{figure}

While the empirical specification in Table~\ref{tab:estimates1} is simple, the behavioral implications of the dynamic model, which we illustrate in Figure~\ref{fig:mtheta}, are quite rich.
Theorem~\ref{th1} shows that $m(X)'\theta$ corresponds to the difference in the choice-specific value functions for quitting and continuing, $m(X)'\theta=V_1(X)-V_0(X)$, at each value of the state variables $X=(hours\ worked, income)$.   Hence, in Figure~\ref{fig:mtheta}, we plot the estimated $m(X)'\theta$ function at different levels of hours worked.  
Clearly, the $m(X)'\theta$ curves are increasing in both income and hours worked (except at hours worked eqaul to ten).   That is, for most values of the state variables $X$, the continuation benefits from quitting, $V_1(X)$, are increasing faster than the continuation benefits from continuing to drive, $V_0(X)$, as income rises and hours worked increases.   This implies that, holding hours worked fixed, drivers are more likely to quit as their income increases; similarly, holding income fixed, drivers are more likely to quit as their hours worked increases.

In Figure~\ref{fig:mtheta}, we also plot, using stars, the actual average income in the raw data earned by drivers who quit at different values of hours worked.  This illustrates the ranges of income for which the $m(X)'\theta$ functions would be estimated most precisely.\footnote{Currently, we do not include standard error bands in Figure~\ref{fig:mtheta} as that would complicate the picture substantially.}

A large part of the existing empirical literature on taxicab driver behavior has focused on testing whether drivers' wage elasticities are positive or negative, where positive elasticities are viewed as a corroboration of the classic model of labor-leisure choice, and negative elasticities are taken as evidence of a behavioral ``income targeting'' model.   In both types of models, the wage rate is taken to be exogenous by the drivers and unchanging throughout the course of the day. Drivers then decide how many hours to work for each given wage rate.  

In our model, however, income evolves stochastically; drivers' ``wage rates'' are random and vary across the shift. Accordingly, a driver reconsiders the decision to end the shift after each fare.
Hence, our modelling framework is not ideal for computing wage elasticities.   That caveat aside, the results in Figure \ref{fig:mtheta} do have implications for wage elasticities if we associate higher cumulative income means with a higher wage rate (as is done in the existing literature).  Under this interpretation, the graphs suggest that by holding hours fixed, the quitting probability is increasing in wage rate, which is consistent with negative wage elasticities.  To illustrate this, consider a thought experiment where we have two drivers who have only driven four hours; the first driver already has income of \$300, while the second only has income of \$180.   The graphs suggest that the benefits from quitting are relatively larger for the first rather than the second driver, and hence that the first should quit more readily than the second.  

On the other hand, if we just use reduced-form evidence from the raw data, the clear upward-sloping trend in the stars suggests the opposite result: that hours worked is increasing in wage (positive income elasticities).  This highlights the benefits of estimating a structural dynamic behavioral model. 
Moreover, our results show that, once dynamics are modelled, it is possible to obtain ``nonstandard'' (i.e., negative) wage elasticities from a model in which drivers' utility functions do not have any explicitly ``nonstandard'' features, such as reference dependence or loss aversion.

\section{Conclusions}
In this paper we consider the estimation of dynamic binary discrete choice models in a semiparametric setting, in which the per-period utility functions are parameterized as single-index functions, but the distribution of the utility shocks is left unspecified and treated as nuisance components of the model. This setup differs from most of the existing work on estimation and identification of dynamic discrete choice models. For identification, we derive a new recursive representation for the unknown quantile function of the utility shocks; our argument requires no additional exclusion restrictions beyond the conditional independence conditions assumed in the typical parametric dynamic-discrete choice literature (e.g. Rust (1987, 1994)). Accordingly, we obtain a single-index representation for the conditional choice probabilities in the model, which permits us to estimate the model using classic estimators from the existing semiparametric binary choice literature.

In particular, we use Powell, Stock and Stoker's (\citeyear{powell1989semiparametric}) kernel-based estimator to estimate the dynamic discrete choice model.   We show that the estimator has the same asymptotic properties as PSS's original estimator (for static discrete-choice models), under mild conditions.   Significantly, the estimator is simple to compute, because it does not require repeated iterations to find a solution.   Monte Carlo simulations show that the estimator works well even in moderately-sized samples.   We provide an empirical application to estimate the dynamic labor supply problem for taxicab drivers in New York City.

More broadly, the analysis in this paper has opened possibilities for the use of classic estimators from the semiparametric literature, which were proposed for estimation of static model, to dynamic models.   We will continue exploring these possibilities in future work.


\bibliographystyle{econometrica}
\bibliography{overall}

\clearpage
\appendix
\small

\section{Proofs} 
\subsection{Proof of \Cref{lemma1}}  \label{prooflemma1}
\proof First, note that the resolvent kernel $R^*$ is well--defined. This is because $\beta^{s-1} f_{X^{[s]}|X}(x'|x)\rightarrow 0$ as $s\rightarrow+\infty$. Under \cref{as_regular}, the solution $V^e(x)$ is also well defined.

Because  it is straightforward to verify that the solution in the lemma solves eq. \eqref{bellman}, Hence, it suffices to show the uniqueness of the solution. 
Eq. \eqref{bellman} can be rewritten as
\[
  V^e(x)=u^e(x)+\beta\cdot \int  V^e(x')\cdot f_{X'|X}(x'|x)d x', \ \ \forall \ x\in\mathscr S_X,
 \]which is an FIE--2.  Then, we apply the method of Successive Approximation \citep[see e.g.][]{zemyan2012classical}. Specifically, let $ V^*(\cdot)$ be an alternative solution to \eqref{bellman}. Then, we have
\[
V^*(x)=u^e(x)+ \beta\int_{\mathscr S_X} V^*(x') \cdot f_{X'|X} (x'|x) dx'.
\]Let $\nu(x)=V^e(x)-V^*(x)$. Then $\nu(x)$ satisfies the following equation:
\[
\nu(x)=\beta\int_{\mathscr S_X} \nu(x') \cdot f_{X'|X} (x'|x) dx'.
\]It suffices to show that $\nu(\cdot)$ has the unique solution: $\nu(x)=0$. To see this, we substitute the left--hand side as an expression of $\nu$ into the integrand:
\[
\nu(x)=\beta^2\int_{\mathscr S_X} \int_{\mathscr S_X} \nu(\tilde x) \cdot f_{X'|X} (\tilde x|x') d\tilde x \cdot f_{X'|X} (x'|x) dx' =\beta^2\int_{\mathscr S_X}  \nu(x') \cdot f_{X^{[2]}|X}(x'|x) dx'.
\]
Repeating this process, then we have: for all $t\geq 1$
\[
\nu(x)=\beta^t\int_{\mathscr S_X} \nu(x') \cdot f_{X^{[t]}|X} (x'|x) dx'.
\]
For the stationary Markov equilibrium, $f_{X^{[t]}|X} (x'|x)$ converges to $f_X(x')$ as $t\rightarrow \infty$. Hence, the right--hand side converges to zero as $t$ goes to infinity. It follows that $\nu(x)=0$ for all $x\in\mathscr S_X$. 
%
\qed

\subsection{Proof of \Cref{lemma4}}  \label{prooflemma4}
\proof
The result follows the Theorem of Successive Approximation \citep[see e.g.][]{zemyan2012classical}. 
\qed

\subsection{Proof of \Cref{th2}}\label{proofthm2}
The estimator is defined in (\ref{pssestimator}).  For the consistency of $\hat \theta$, we need $h_\theta\rightarrow 0$, $Th_\theta^{k_\theta+1}\rightarrow \infty$ and $\mathbb E |\hat m(X)-m(X)|=o(h_\theta)$ as $T\rightarrow \infty$. Note that the last condition ensures the estimation error in $\hat m$ is negligible. 

Let $\tilde\theta$ be the infeasible estimator 
$$
\tilde \theta =-\frac{2}{T(T-1)}\sum_{t=1}^T \sum_{s\neq t}\left[\frac{1}{h^{k_\theta+1}_\theta}\times \nabla K_\theta\left(\frac{m(X_t)- m(X_s)}{h_\theta}\right)\times Y_s\right].
$$
The asymptotic analysis for $\tilde\theta$ was done in \cite{powell1989semiparametric}.  They show that the variance term in $\tilde \theta$ has the order $T^{-1}$ if $Th_\theta^{k_\theta+2}\rightarrow \infty$, while the bias term has the order $h_\theta^P$. Therefore, if $T^{1/2} h_\theta^p\rightarrow 0$, then the bias term disappear faster than $T^{-1/2}$. The leading term left is the variance term -- the $\tilde \theta$ converges at the rate $T^{-1/2}$.   Our arguments piggybacks off of this argument, as we will show here that $T^{1/2}(\hat\theta-\theta)$ is identical to $T^{1/2}(\tilde\theta-\theta)$ by a negligible factor; that is, our estimator and the infeasible estimator have the same limiting distribution (corresponding to that derived in \cite{powell1989semiparametric}).

%

By Taylor expansion, we have
\begin{multline}
\hat \theta =\tilde \theta -\frac{2}{T(T-1)}\sum_{t=1}^T \sum_{s\neq t}\left[\frac{1}{h^{k_\theta+2}_\theta} \nabla^2 K_\theta\left(\frac{ m(X_t)- m(X_s)}{h_\theta}\right) \times Y_s\times \big(\hat m(X_t)- m(X_t) \big)\right]\\
+ \frac{2}{T(T-1)}\sum_{t=1}^T \sum_{s\neq t}\left[\frac{1}{h^{k_\theta+2}_\theta} \nabla^2 K_\theta\left(\frac{ m(X_t)- m(X_s)}{h_\theta}\right) \times Y_s\times \big(\hat m(X_s)- m(X_s) \big)\right]\\
+O_p(h_\theta^{-3}\cdot \mathbb E \|\hat m(X)-m(X)\|^2)
\equiv \tilde \theta + \mathbb A_1 + \mathbb A_2 + \mathbb B
\end{multline}

 We will show that $\mathbb A_1 + \mathbb A_2 + \mathbb B_2$ are all $o_p(T^{-1/2})$ implying $T^{1/2} (\hat\theta-\tilde\theta)$ is negligible.
First, by  \Cref{as8}(i), we have 
\begin{equation}
\label{eq2a}
h_\theta^{-3}\times \mathbb E \|\hat m(X)-m(X)\|^2= h_\theta^{-3}\times o_p(T^{-1/2}h_\theta^3)=o_p(T^{-1/2}).
\end{equation} Then, $\mathbb B=o_p (T^{-1/2})$.

Next we show $\mathbb A_1$ and $\mathbb A_2=o_p(T^{-1/2})$. For simplicity, we only provide an argument for  $\mathbb A_1$ (that for $\mathbb A_2$ is analogous).

Define
$$
\tilde {\mathbb A}_1\equiv- \frac{2}{T(T-1)}\sum_{t=1}^T \sum_{s\neq t}\left[\frac{1}{h^{k_\theta+2}_\theta} \nabla^2 K_\theta\left(\frac{ m(X_t)- m(X_s)}{h_\theta}\right)  Y_s\times \big[\mathbb E[\hat m(X_t)|X_t,X_s]- m(X_t) \big]\right]
$$
Clearly $\mathbb E (\mathbb A_1)=\mathbb E(\tilde {\mathbb A}_1)$.  Following \cite{powell1989semiparametric}, we now establish two properties:
\begin{align*}
&(a): \tilde {\mathbb A}_1 = o_p(T^{-1/2});\\
&(b): T\times \text{Var} ( \mathbb A_1 -\tilde {\mathbb A}_1)\rightarrow 0,
\end{align*} 
which together imply $ \mathbb A_1=o_p(T^{-1/2}) $.

For property (a), by \Cref{as8}(iii), 
\[
\mathbb E[\hat m(X_t)|X_t,X_s]=\mathbb E[\hat m_{t,-s}|X_t,X_s]+o_p(T^{-1/2}h_\theta^2)=\mathbb E[\hat m(X_t)|X_t]+o_p(T^{-1/2}h_\theta^2).
\] Then, we have
\begin{eqnarray*}
&&\tilde {\mathbb A}_1\\
&=&- \frac{2}{T(T-1)}\sum_{t=1}^T \sum_{s\neq t}\left[\frac{1}{h^{k_\theta+2}_\theta} \nabla^2 K_\theta\left(\frac{ m(X_t)- m(X_s)}{h_\theta}\right)  Y_s\times \big[\mathbb E[\hat m(X_t)|X_t]- m(X_t) \big]\right]+ o_p(T^{-1/2})\\
&\equiv& \mathbb C_1 + o_p(T^{-1/2}).
\end{eqnarray*}
Because 
\begin{eqnarray*}
\mathbb E |\mathbb C_1|&\leq& 2\mathbb E \left|\frac{1}{h^{k_\theta+2}_\theta} \nabla^2 K_\theta\left(\frac{ m(X_t)- m(X_s)}{h_\theta}\right)  \times \big[\mathbb E[\hat m(X_t)|X_t]- m(X_t) \big]\right|\\
&\leq &2\overline C\times \frac{1}{h^{2}_\theta} \mathbb E   \left\|\mathbb E[\hat m(X)-m(X)|X] \right\|
\end{eqnarray*}for some positive $\overline C<\infty$.
 Hence, by \Cref{as8}(ii), property (a) obtains.

For property (b), note that
\begin{eqnarray*}
\mathbb A_1-\tilde{\mathbb A}_1&\equiv &- \frac{2}{T(T-1)}\sum_{t=1}^T \sum_{s\neq t} \phi_{T,s,t}\times \big[\hat m(X_t) -\mathbb E[\hat m(X_t)|X_t] \big]+o_p(T^{-1/2})\equiv \mathbb C_2+o_p(T^{-1/2})
\end{eqnarray*}where $\phi_{T,s,t}=\frac{1}{h^{k_\theta+2}_\theta} \nabla^2 K_\theta\left(\frac{ m(X_t)- m(X_s)}{h_\theta}\right)  Y_s$.

Clearly, 
\begin{eqnarray*}
&&\text{Var} (\mathbb C_2)=\frac{4}{T^2(T-1)^2}\sum_{t=1}^T \sum_{s\neq t} \text{Var}\Big(\phi_{T,s,t}\times \big[\hat m(X_t) -\mathbb E[\hat m(X_t)|X_t] \big]\Big)\\
&+& \frac{4}{T^2(T-1)^2}\sum_{t=1}^T \sum_{s\neq t}\sum_{s'\neq t,s} \text{Cov}\Big(\phi_{T,s,t}\big[\hat m(X_t) -\mathbb E[\hat m(X_t)|X_t] \big], \phi_{T,s',t} \big[\hat m(X_t) -\mathbb E[\hat m(X_t)|X_t] \big]\Big)\\
&+& \frac{4}{T^2(T-1)^2}\sum_{t=1}^T \sum_{s\neq t}\sum_{t'\neq t,s}\sum_{s'\neq t,s,t'} \text{Cov}\Big(\phi_{T,s,t}\big[\hat m(X_t) -\mathbb E[\hat m(X_t)|X_t] \big], \phi_{T,s',t'} \big[\hat m(X_{t'}) -\mathbb E[\hat m(X_{t'})|X_{t'}] \big]\Big)\\
&=&O(T^{-2}h_\theta^{-k_\theta-4})\times  \mathbb E \big\{\hat m(X) -\mathbb E[\hat m(X)|X] \big\}^2\\
&+& \frac{4}{T}\ \text{Cov}\Big(\phi_{T,2,1}\big[\hat m(X_1) -\mathbb E[\hat m(X_1)|X_1] \big], \phi_{T,3,1} \big[\hat m(X_1) -\mathbb E[\hat m(X_1)|X_1] \big]\Big)\\
&+& 4\ \text{Cov}\Big(\phi_{T,2,1}\big[\hat m(X_1) -\mathbb E[\hat m(X_1)|X_1] \big], \phi_{T,4,3} \big[\hat m(X_3) -\mathbb E[\hat m(X_3)|X_3] \big]\Big).
\end{eqnarray*}
Note that 
\begin{eqnarray*}
&&\text{Cov}\Big(\phi_{T,2,1}\big[\hat m(X_1) -\mathbb E[\hat m(X_1)|X_1] \big], \phi_{T,4,3} \big[\hat m(X_3) -\mathbb E[\hat m(X_3)|X_3] \big]\Big)\\
&=&\mathbb E \left\{\phi_{T,2,1}\phi_{T,4,3} \big[\hat m(X_1) -\mathbb E[\hat m(X_1)|X_1] \big] \times \big[\hat m(X_3) -\mathbb E[\hat m(X_3)|X_3] \big]\right\}\\
&-&\mathbb E \left\{\phi_{T,2,1} \big[\hat m(X_1) -\mathbb E[\hat m(X_1)|X_1] \big] \right\}\times \mathbb E \left\{\phi_{T,4,3} \big[\hat m(X_3) -\mathbb E[\hat m(X_3)|X_3] \big]\right\}.
\end{eqnarray*}
By \Cref{as8}(iii), 
\begin{multline*}
\mathbb E \left\{\phi_{T,2,1} \big[\hat m(X_1) -\mathbb E[\hat m(X_1)|X_1] \big] \right\}\\
=\mathbb E \left\{\phi_{T,2,1} \big[\hat m_{1,-2} -\mathbb E[\hat m_{1,-2}|X_1] \big] \right\}+ O_p(h^{-2}_\theta)\times o_p(T^{-1/2}h^2_\theta)=o_p(T^{-1/2}).
\end{multline*}
Furthermore, by the law of iterated expectation (conditioning on the sigma algebra: $\mathscr F_2$,$\mathscr F_4$, $\mathscr F_{5,\cdots,n}$), 
\begin{eqnarray*}
&&\mathbb E \left\{\phi_{T,2,1}\phi_{T,4,3} \big[\hat m(X_1) -\mathbb E[\hat m(X_1)|X_1] \big] \times \big[\hat m(X_3) -\mathbb E[\hat m(X_3)|X_3] \big]\right\}\\
&=&O_p(h^{-4}_\theta)\times o_p(T^{-1/2}h^2_\theta)\times o_p(T^{-1/2}h^2_\theta)\\
&=&o_p(T^{-1}),
\end{eqnarray*}
where the term $o_p(T^{-1/2}h^2_\theta)$ is due to the differences $\hat m(X_1)-\hat m_{1,-3}$ and $\hat m(X_3)-\hat m_{3,-1}$. Therefore, the last term in $\text{Var}(\mathbb C_2)$ is $o_p(T^{-1})$.

Moreover, because 
\begin{eqnarray*}
&& \frac{1}{T}\text{Cov}\Big(\phi_{T,2,1}\big[\hat m(X_1) -\mathbb E[\hat m(X_1)|X_1] \big], \phi_{T,3,1} \big[\hat m(X_1) -\mathbb E[\hat m(X_1)|X_1] \big]\Big)\\
&= &\frac{1}{T}\mathbb E \left\{\phi_{T,2,1}\phi_{T,3,1} \big[\hat m(X_1) -\mathbb E[\hat m(X_1)|X_1] \big]^2\right\}\\
&=&O(T^{-1}h_\theta^{-4})\times    \mathbb E \left\{ \hat m(X)- \mathbb E [\hat m(X)|X]\right\}^2.
\end{eqnarray*}
Then a sufficient condition for property (b) is 
$$
  \mathbb E \left\{ \hat m(X)- \mathbb E [\hat m(X)|X]\right\}^2=o(h^{4}_\theta). 
$$ Note that this condition is implied by \Cref{as8}(i).

 Hence, we have shown that our estimator $\hat\theta$ and the infeasible estimator $\tilde{\theta}$ differ by an amount which is $o_p(T^{-1/2})$.  Hence, the asymptotic properties for $\hat\theta$ are the same as those for the infeasible estimator $\tilde\theta$, which were previously established in \cite{powell1989semiparametric}.
\end{document}